\documentclass[aps,prl,twocolumn,superscriptaddress,amssymb]{revtex4}
\usepackage{graphicx}
\usepackage{amsmath}
\usepackage{amsfonts}
\usepackage{epsfig}
\usepackage{bm}
\usepackage{color}

\begin{document}

\title{Superconductivity in the doped $t$-$J$ model:  results for four-leg cylinders}
\author{Hong-Chen Jiang} 
\email{hcjiang@stanford.edu}
\affiliation{Stanford Institute for Materials and Energy Sciences, SLAC and Stanford University, Menlo Park, California 94025, USA}
\author{Zheng-Yu Weng}
\affiliation{Institute for Advanced Study, Tsinghua University, Beijing, China}
\author{Steven A. Kivelson}
\affiliation{Physics Department, Stanford University, Stanford, CA, 94305, USA}

\date{\today}

%%=============Abstract============
\begin{abstract}
We report a density-matrix renormalization group study of the lightly doped $t$-$J$ model on a 4-leg cylinder with doped hole concentrations per site $\delta=5\%\sim 12.5\%$. By keeping an unusually large number of states and long system sizes, we are able to accurately document the interplay between d-wave superconductivity (SC) and spin and charge-density-wave (CDW) orders.  The long-distance behavior is consistent with that of a Luther-Emery liquid with a spin-gap, power-law  CDW correlations with wave-length $\lambda =1/2\delta$, and power-law SC correlations.  This is the widest (most nearly 2D) such system for which power law SC correlations have been established.
\end{abstract}
\maketitle

%%============Introduction=============
%\textbf{Introduction:} %
The Hubbard model, and the closely related $t$-$J$ model, play central roles in the theory of highly correlated electronic systems.  Enormous effort has been devoted to studying the properties of these models at intermediate couplings.  No general theoretically controlled methods exist for this class of problem.  However, it is possible to obtain essentially exact results on long but moderately narrow ladders and cylinders using density matrix renormalization group (DMRG) methods.\cite{White_PRL_1992}   Cylinders have the local lattice geometry of the two-dimensional (2D) system, and can be  extrapolated to infinite length, {\it i.e.} the thermodynamic limit can be taken in one direction.  Thus, one can hope to obtain insight into the nature of the 2D problem from these solutions.

In this paper, we report extensive DMRG studies of the 4-leg $t$-$J$ cylinder, keeping large numbers of states so that subtle long-distance correlations can be reliably studied.  In addition to the hope that they may shed light on the 2D problem, there are two other reasons to engage in such studies.  Firstly, there is interesting physics of multicomponent one-dimensional (1D) systems that can be directly explored without undue speculation - the only extrapolations are to the limits of zero truncation error and  infinite system length.  Secondly, these systems can be used to benchmark less clearly justified but more widely applicable computational methods.

\begin{figure}
  \includegraphics[width=\linewidth]{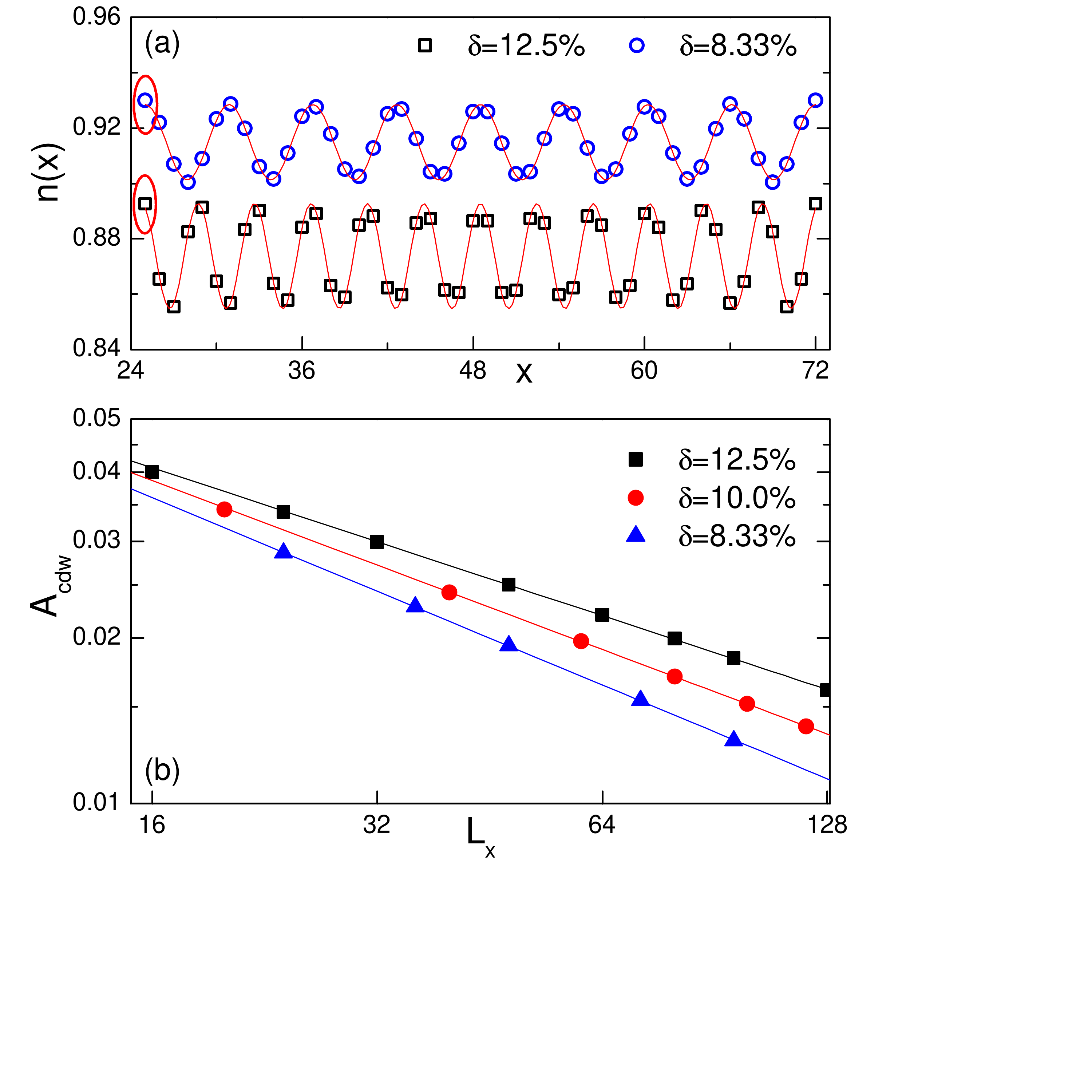}
  \caption{(Color online) (a) Charge density profile $n(x)$ of the $t$-$J$ model at doping levels $\delta=8.33\%$ and $\delta=12.5\%$ on a $L_x=96$ cylinder. The open squares and circles denote numerical data, while the red lines are fits to $n(x)=A_{cdw} \cos (Q x + \theta) +n_0$, where $A_{cdw}$ and $Q$ are the amplitude  and ordering wavevector of the local CDW, respectively. Note that only the central-half region with rung indices $\frac{L_x}{4}< x \leq  \frac{3L_x}{4}$ are shown and used in the fitting while the remaining  $\frac{L_x}{4}$ data points from each end are removed to minimize the boundary effect. The red oval labels the ``reference site'' chosen to calculate the SC correlation function in Eq. (\ref{Eq:SCOR}). (b) Finite-size scaling of CDW amplitude $A_{cdw}$ as a function of $L_x$ at different doping levels $\delta$ in a double-logarithmic plot.}\label{Fig:CDWP}
\end{figure}

\begin{figure}
  \includegraphics[width=\linewidth]{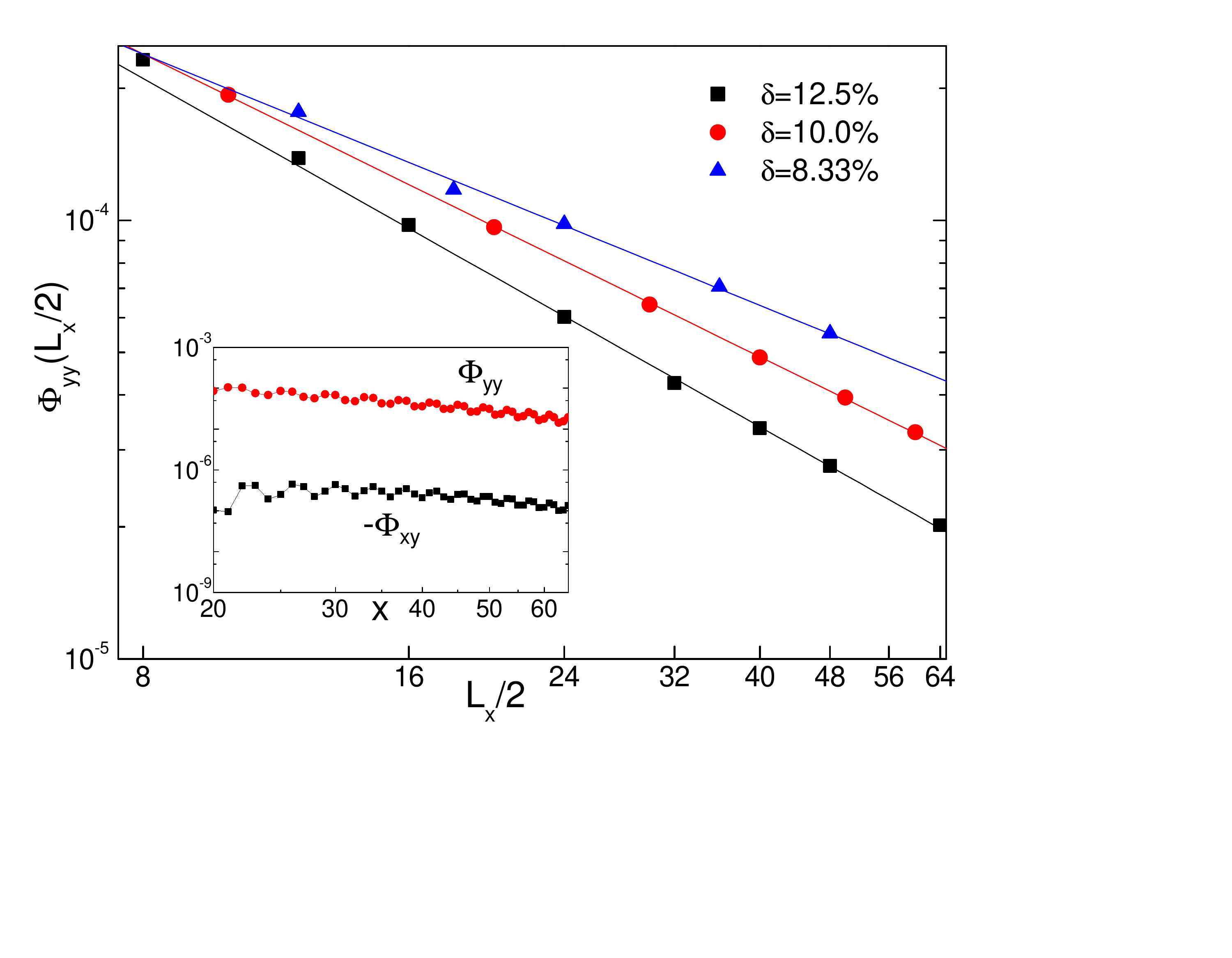}
  \caption{(Color online) Finite-size scaling of superconducting correlation $\Phi_{yy}(\frac{L_x}{2})$ as a function of $L_x$ in a double-logarithmic plot at different doping levels on a log-log plot. The solid lines are fits to $\Phi_{yy}(\frac{L_x}{2})\sim (\frac{L_x}{2})^{-K_{sc}}$. Inset: Superconducting pair-field correlations $\Phi_{yy}$ and -$\Phi_{xy}$ on $L_x=128$ cylinder as a function of displacement along $\hat{x}$ direction.
  } \label{Fig:SCP}
\end{figure}

\textbf{Principal Findings:} We have studied the equal time superconducting (SC), charge-density wave (CDW) and spin-density wave (SDW) correlations for a range of doped hole concentrations, $\delta= 5$\%, 8.33\%,  10\%, and 12.5\%, and for a characteristic value of $t/J =3$.  (Representative results are shown in Figs. \ref{Fig:CDWP}, \ref{Fig:SCP}, \ref{Fig:SDWP}, and \ref{Fig:ABC}.  We have obtained similar, but less extensive results for other values of $t/J$.)  Thought of as  a 1D system, we find that the ground-state is always in a Luther-Emery (LE) phase,\cite{lutheremery} characterized by a finite spin-gap as a consequence of exponential decay of spin correlations, and CDW and SC correlations that fall at long distances as $\cos(Qr+\theta) \  r^{-K_c}$ and $r^{-K_{sc}}$ respectively, where the CDW wave-vector $Q=4\pi \delta$.  This is consistent with recent study of $t$-$t^\prime$-$U$ Hubbard model at doped hole concentration $\delta=12.5\%$ on 4-leg cylinders.\cite{Jiang_Hubbard_4leg}

\begin{figure}
  \includegraphics[width=\linewidth]{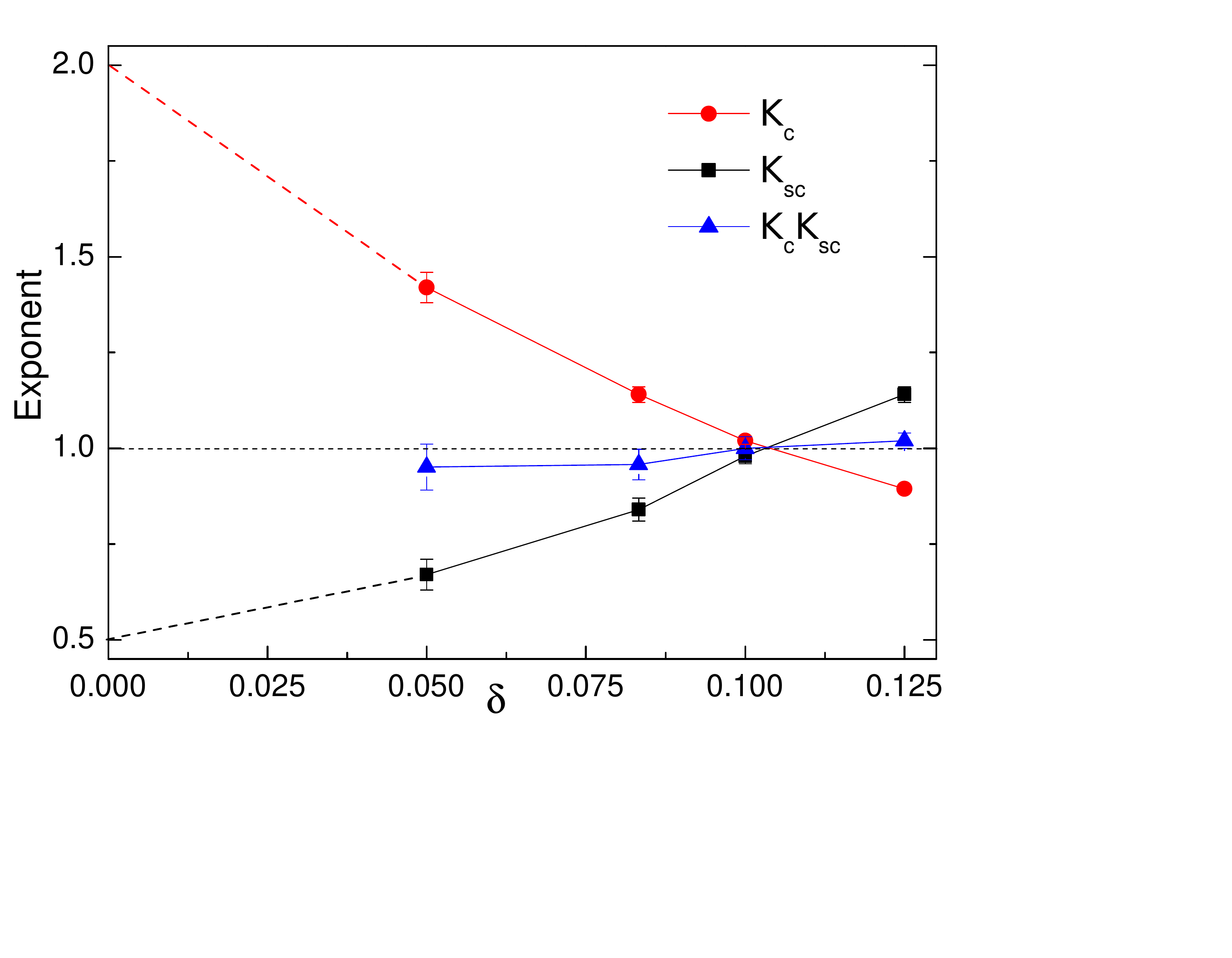}
  \caption{(Color online) Luttinger exponents $K_{c}$, $K_{sc}$ and their product $K_{c}K_{sc}$, as a function of doping level $\delta$. The filled symbols represent the extracted values from the fits in Fig.\ref{Fig:CDWP}(b) and Fig.\ref{Fig:SCP}. The solid and dotted lines are guides to the eye.}\label{Fig:LEL}
\end{figure}

\begin{figure}
  \includegraphics[width=\linewidth]{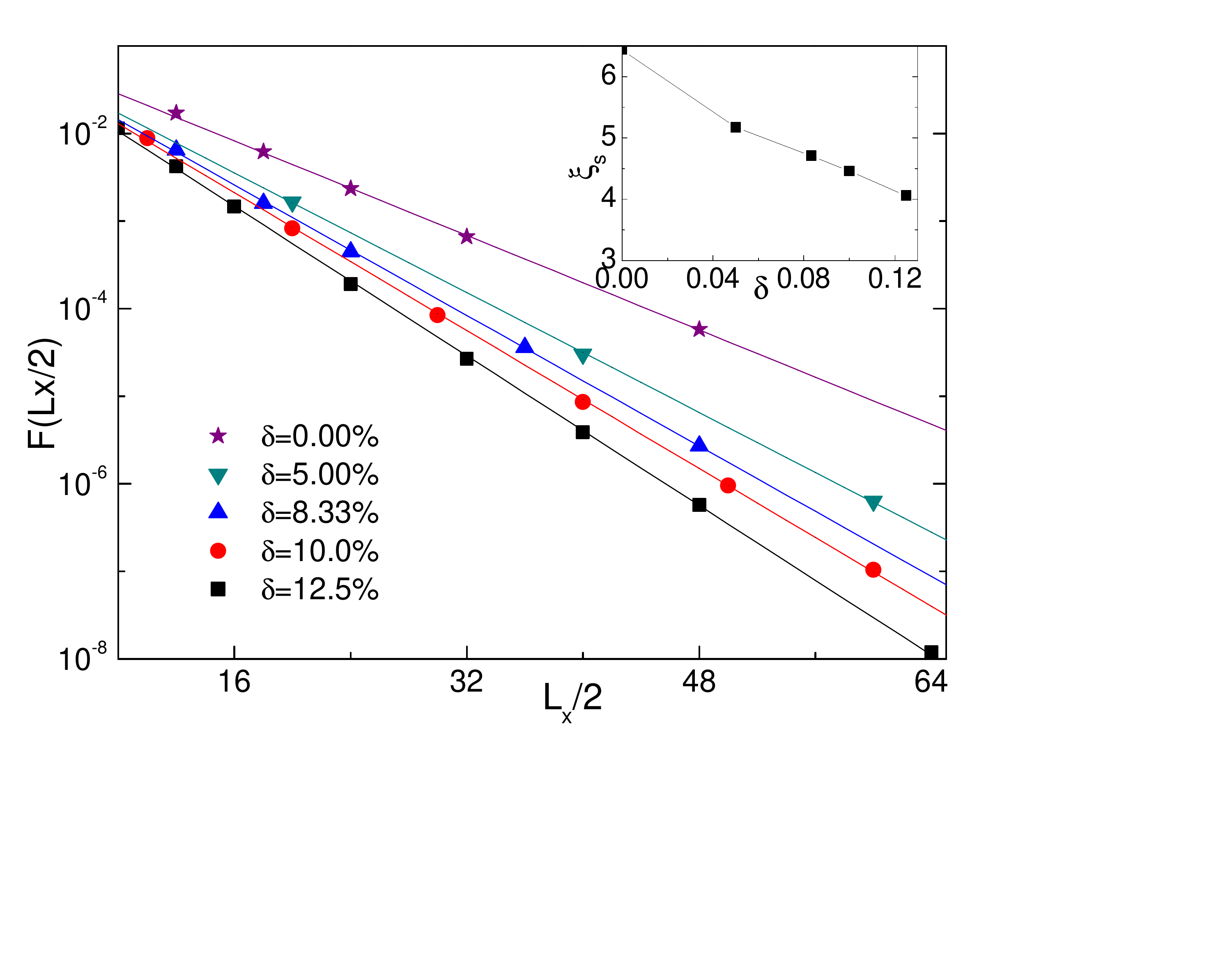}
  \caption{(Color online) Finite-size scaling of the spin correlation function $F(\frac{L_x}{2})$ as a function of $L_x$ at doping levels $\delta=0\%\sim 12.5\%$ in the semi-logarithemic plot. The inset shows the correlation length $\xi_s$ obtained from fits (solid lines) to data in the main panel using $F(\frac{L_x}{2})\propto e^{-L_x/2\xi_s}$.}\label{Fig:SDWP}
\end{figure}

\begin{figure}
  \includegraphics[width=\linewidth]{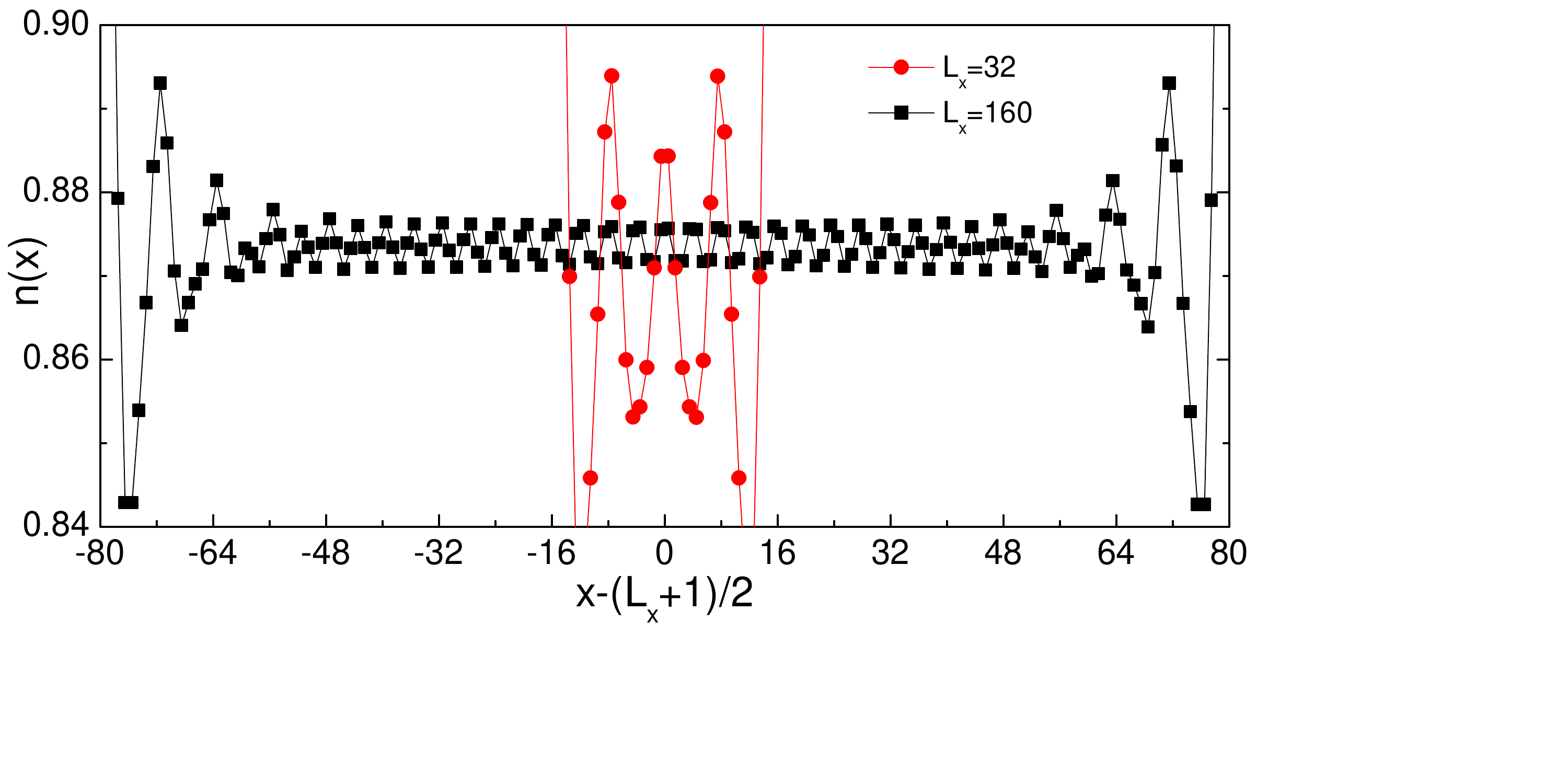}
  \caption{(Color online) The charge density profile $n(x)$ for the $t$-$J$ model %at doping level
  with $\delta=12.5\%$ and  anti-periodic boundary conditions in the $\hat{y}$ direction for $L_x=32$ and $L_x=160$ cylinders.}\label{Fig:ABC}
\end{figure}

\begin{figure}
  \includegraphics[width=\linewidth]{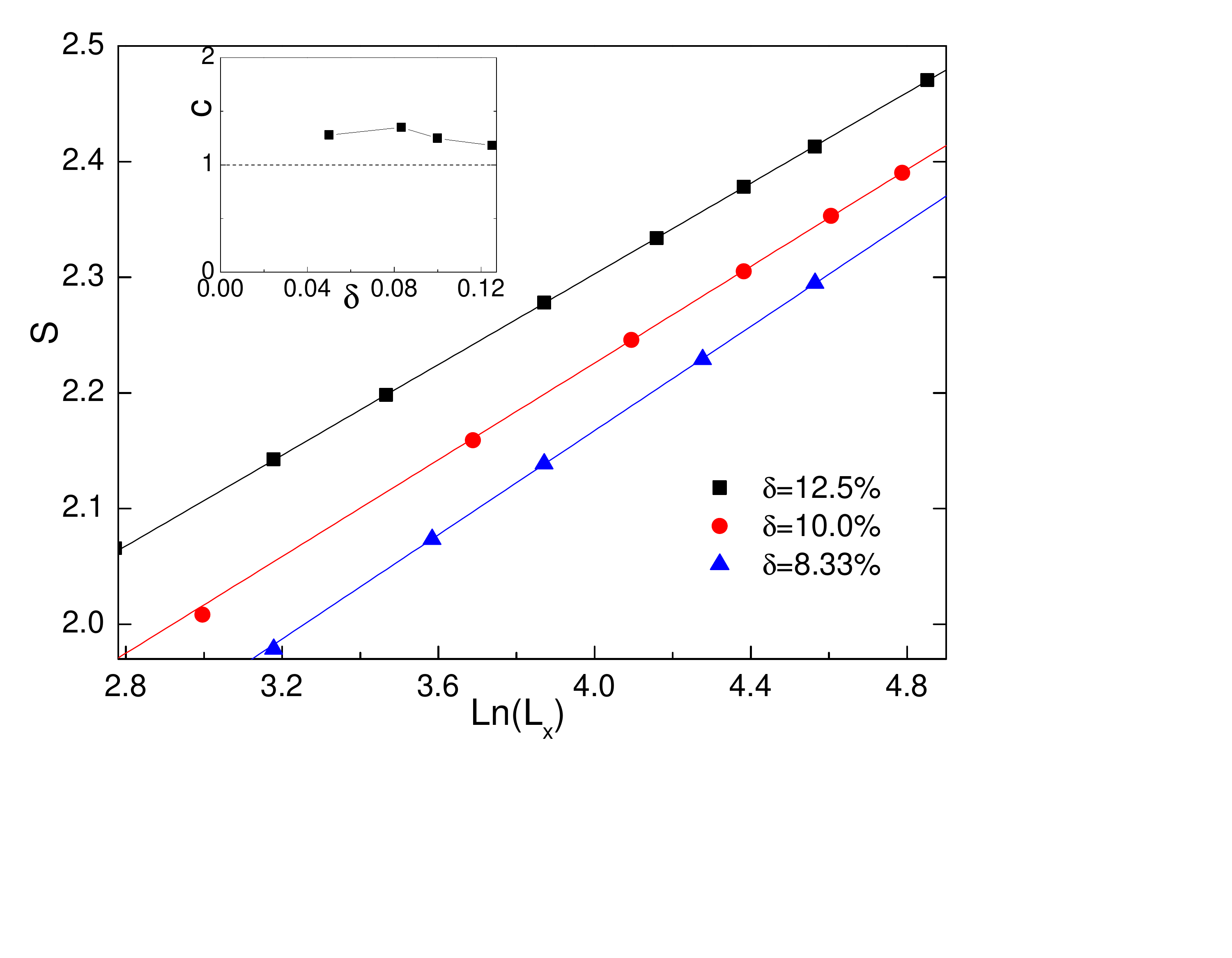}
  \caption{(Color online) Von Neumann entanglement entropy $S$ for the $t$-$J$ model with $\delta=8.33\%$, $10\%$ and $12.5\%$. Inset: The extracted central charge $c$ as a function of $\delta$. Dashed line marks $c=1$.
  }\label{Fig:Entropy}
\end{figure}

(An ordered state with this value of $Q$ has wave-length $\lambda=1/2\delta$, and so half a doped hole per unit cell corresponding to what is referred to as ``half-filled'' stripes.)  Moreover, within numerical uncertainty, as theoretically expected of a LE liquid, $K_c K_{sc} =1$ (Fig. \ref{Fig:LEL}) and the central charge, $c$, extracted from the scaling of the entanglement entropy, is $c=1$  (inset of Fig. \ref{Fig:Entropy}). The SC and CDW correlations are invariant with respect to the $C_4$ symmetry of rotations about the axis of the cylinder.   The SC correlations have a ``d-wave-like'' form factor in that the sign of the pair-field is opposite on  bonds  perpendicular to  and along the cylinder (Y-directed and X-directed bonds).  However, this is far from a statement of symmetry, and indeed there is an almost equal in strength admixture of an ``extended s-wave'' component with the consequence that  the pair field  amplitude on Y-bonds is two orders of magnitude larger than on X-bonds.  (See Fig.\ref{Fig:SCP})

For all the dopings studied,  $K_{sc}<2$ and $K_c<2$, which  (assuming  the usual emergent Lorentz invariance)  implies that both the corresponding susceptibilities  diverge as $T\to 0$, as $T^{-(2-K_{sc})}$ and $T^{-(2-K_c)}$ respectively. As far as we know, this is the first demonstration of power-law superconducting correlations (and by implication a divergent SC susceptibility) on such a wide ladder or cylinder.  As shown in Fig. \ref{Fig:LEL}, $K_{sc}$ is an increasing function of $\delta$ and $K_c$ a decreasing function, so that the SC susceptibility is more divergent for $\delta < 0.1$ and the CDW is more divergent for $\delta > 0.1$.

In a previous study\cite{Dodaro_PRB_2017}, we explored the same model over a wider range of parameters (including both first and second neighbor couplings, $t^\prime$ and $J^\prime$). The primary focus of this earlier study was to explore the extent to which the nature of the ground-state  (especially the nature of the CDW order) depends on ``microscopic details.''  For the special case $t^\prime = J^\prime = 0$, on which we focus here, these earlier results are  generally consistent with our present results.  However,   the longer system sizes (up to $L_x = 128$) and the larger number of states used in the present study increase our ability to distinguish exponentially falling correlations, power law (quasi-long-range) order and true long-range order.   In particular, what was previously tentatively identified as SC order with a long but finite correlation length, we now identify as quasi-long-ranged SC order, albeit with exactly the previously determined form factor.

%===============Fig1: Charge density wave===============

%%===============Fig2: Superconducting correlation===============

%%=========Model Hamiltonian==============
\textbf{Model and Method:}
We study the hole-doped $t$-$J$ model on the square lattice defined by the Hamiltonian%
\begin{eqnarray}\label{Eq:ModelHamiltonian}
H =-t \sum_{\langle ij\rangle \sigma} \left(\hat{c}^+_{i\sigma} \hat{c}_{j\sigma} + h.c.\right)+ J\sum_{\langle ij\rangle} \left ( \vec{S}_i\cdot \vec{S}_j - \frac{\hat{n}_i\hat{n}_j}{4}\right ), 
\end{eqnarray}
where $\hat{c}^+_{i\sigma}$ ($\hat{c}_{i\sigma}$) is the electron creation (annihilation) operator on site $i=(x_i,y_i)$ with spin $\sigma$, $\vec{S}_i$ is the spin operator and $\hat{n}_i=\sum_{\sigma}\hat{c}^+_{i\sigma}\hat{c}_{i\sigma}$ is the electron number operator, $\langle ij\rangle$ denotes nearest-neighbor (NN) sites and the Hilbert space is constrained by the no-double occupancy condition $\hat{n}_i\leq 1$.  The parameters $t$ and $J$ are the electron hopping integral and   the spin superexchange interactions between NN sites. We take the lattice geometry to be cylindrical and a lattice spacing of unity. Thus, unless stated otherwise, we take periodic boundary conditions in the $\hat{y}=(0,1)$ direction and open in the $\hat{x}=(1,0)$ direction, although for comparison we also consider the case of anti-periodic boundary conditions corresponding to a half-quantum of flux threaded along the axis of the cylinder.  Here, we focus on cylinders with circumference $L_y=4$ and length $L_x$. There are $N=L_x\times L_y$ lattice sites and the number of electrons is $N_e=N$ at half-filling, i.e., $\hat{n}_i=1$, where the $t$-$J$ model reduces to the spin-$\frac{1}{2}$ antiferromagnetic Heisenberg model. The concentration of ``doped holes'' is defined as $\delta=\frac{N_h}{N}$, where $N_h=N-N_e$. 

For the present study, we focus on the lightly doped case at doping levels in the range $\delta=5\%\sim 12.5\%$ on cylinders with length up to $L_x=128$. We set $J=1$ as the energy unit and report results for $t=3$. 
We keep the total magnetization fixed at zero and perform around 60 sweeps and keep up to $m=15000$ states in each DMRG block with a typical truncation error $\epsilon\lesssim 1\times 10^{-7}$ for all doping levels. This leads to excellent convergence for our results when extrapolated to $m=\infty$ limit. In all cases, but especially when computing SC correlations, it proves essential to keep very large $m$ and to analyze the $m\to \infty$ seriously, and in some cases, in particular in the case in which there is a half-flux quantum along the axis of the cylinder, it is necessary to go to system sizes much longer than $L_x=48$ in order to observe the correlations that arise in the $L_x\to \infty$ limit. Further details concerning what is required to establish convergence to the ground-state in each case is presented in the Supplemental Material.

%===============Fig3: Luther-Emery liquid===============

%%===========Theoretical expectation============
\textbf{Theoretical expectations:}
In a Luther-Emery liquid phase, there is a single gapless spinless bosonic mode with linear dispersion (emergent Lorentz symmetry) - {\it i.e.} it is asymptotically equivalent to a 1+1 dimensional conformal field theory (CFT) with $c=1$.  At  long-distances the density-density correlation function oscillates with a well-defined wave-vector $Q$ and decays with a power law given by the Luttinger exponent, $K_c$, while the dual SC correlations exhibit non-oscillatory power law decay with exponent $K_{sc}=1/K_c$.  Because there is a spin-gap,  spin correlations fall exponentially with a finite correlation length $\xi_s$;  if $\xi_s$ is sufficiently large, one can also frequently identify a wave-vector $Q_{sdw}$ which characterizes the oscillations of the SDW correlations.  

These properties can be extracted in various ways from numerical data.  Because the CDW is pinned by the cylinder ends, an  effective method to study the CDW correlations is to compute the charge density modulations in the middle region of a long finite cylinder, $\langle \hat n_i\rangle \approx (1-\delta) + A_{cdw}(L_x)\  \cos(Q x_i + \theta)$ for $x_i$ near $L_x/2$.
The SC correlations are determined from the long distance behavior of the modulus of the SC correlator, $\Phi_{\alpha,\beta}(x)$ defined in Eq.(\ref{Eq:SCOR}).  The expectation is that  the decay of these quantities  is governed by the appropriate exponents,
\begin{eqnarray}
A_{cdw}(L_x)\propto L_x^{-K_c/2}  \ {\rm and} \ %
\Phi_{\alpha\beta}(x)\propto |x|^{-K_{sc}}, \label{Eq:Kc}
\end{eqnarray}
where the second relation applies for displacements along the cylinder $1\ll |x| \ll L_x$.  Similarly, $Q_{sdw}$ and $\xi_{sdw}$ can be extracted from the long-distance behavior fo the spin-spin correlation function.

%%==========CDW correlations=============
\textbf{CDW correlations:} To describe the charge density properties of the system, we define the local rung density operator as $\hat{n}(x)=\frac{1}{L_y}\sum_{y=1}^{L_y}\hat{n}(x,y)$ and its expectation value as $n(x)=\langle \hat{n}(x)\rangle$. Fig. \ref{Fig:CDWP} shows $n(x)$ in a central portion of cylinders with $L_x=96$ for $\delta=8.33\%$ and $\delta=12.5\%$. Here, a stripe pattern with wavelength $\lambda=1/2\delta$ is found, {\it i.e.} $\lambda=4$ for $\delta=12.5\%$,  consistent with previous studies.\cite{Siller_PRB_2002,Dodaro_PRB_2017}.
Similar behavior (not shown) is found at other doping levels. That the oscillations appear not to decrease with increasing distance from the cylinder ends might be taken as evidence of long-range CDW order (possibly due to a commensurability lock-in).

However, the amplitude of the CDW decreases with increasing $L_x$.  Fig.\ref{Fig:CDWP}(b) shows examples of finite-size scaling of $A_{cdw}$ as a  function of $L_x$. In the double-logarithmic plot, our results for all doping levels are approximately linear, which suggests that $A_{cdw}(L_x)$ decays with a power-law and vanishes as $L_x\to \infty$. The exponent $K_c$, which is shown in Fig.\ref{Fig:LEL}, was obtained by fitting the data points using Eq. (\ref{Eq:Kc}). $K_c$ can also be obtained directly from the decay of the  density-density correlation function near the cylinder ends (see the Supplemental Material).

%%=========Superconducting correlation===============
\textbf{Superconducting correlation:} %
Since the ground state of the system with any even number of doped holes is always found to have spin 0, we will focus on spin-singlet pairing. A diagnostic of the superconducting order is the pair-field correlator defined as%
\begin{eqnarray}
\Phi_{\alpha\beta}(x)=\frac{1}{L_y}\sum_{y=1}^{L_y}\ \langle \Delta_\alpha^\dagger (x_0,y)\Delta_\beta(x_0+x,y)\rangle.\label{Eq:SCOR}
\end{eqnarray}
Here the spin-singlet pair-field creation operator is $\Delta_\alpha^\dagger(x,y)=\frac{1}{\sqrt{2}}[c_{(x,y),\uparrow}^\dagger c_{(x,y)+\alpha,\downarrow}^\dagger - c_{(x,y),\downarrow}^\dagger c_{(x,y)+\alpha,\uparrow}^\dagger]$, where bond orientations are designated $\alpha=\hat{x}$, $\hat{y}$, $(x_0,y)$ is the reference bond indicated by the red oval  shown in Fig.\ref{Fig:CDWP}, and $x$ is the displacement in the $\hat{x}=(1,0)$ direction.

Fig.\ref{Fig:SCP} shows the finite-size scaling of $\Phi_{yy}(\frac{L_x}{2})$ at different  doping levels. It decays with a power-law, whose exponent $K_{sc}$,  plotted in Fig.\ref{Fig:LEL}, was obtained by fitting the results using Eq.(\ref{Eq:Kc}). Therefore, we can conclude that the lightly doped $t$-$J$ model on width $L_y=4$ cylinders has quasi-long-range superconducting correlation. It is worth to note that $K_{sc}$ decreases with decreasing $\delta$ tending to saturate at $K_{sc}=0.5$, while $K_c$ increases and tends to saturate at $K_{c}=2$ as $\delta \to 0$.  Both tendencies are consistent with theoretical prediction.\cite{Balents_PRB_1996}

%========Fig4: Spin-spin correlation function==========

%%==========Spin-spin correlation=============
\textbf{Spin-spin correlation:} %
To describe the magnetic properties of the ground state, we have also calculated the spin-spin correlation functions defined as%
\begin{eqnarray}\label{Eq:SpinCor}
F(x)=\frac{1}{L_y}\sum_{y=1}^{L_y}|\langle \vec{S}_{x_0,y}\cdot \vec{S}_{x_0+x,y}\rangle|.
\end{eqnarray}
Here $\vec{S}_{x,y}$ denotes the spin operator on site $i=(x,y)$. $(x_0,y)$ is the reference site indicated by the red oval  shown in Fig.\ref{Fig:CDWP}, and $x$ is the displacement in the $\hat{x}=(1,0)$ direction. As we did for $A_{cdw}$ and $\Phi_{yy}$, we first extrapolate the spin correlation function $F(\frac{L_x}{2})$ to the limit $m=\infty$, and then analyze the functional dependence of the result on $L_x$. As shown in Fig.\ref{Fig:SDWP}, $F(\frac{L_x}{2})$ decays exponentially with $L_x$, i.e., $F(\frac{L_x}{2})\propto e^{-L_x/2\xi_s}$. The corresponding spin-spin correlation length is $\xi_s= 4\sim 5$ lattice spacings. We conclude that the spin correlations are short-ranged and consequently that there is a spin-gap.

%========Fig5: CDW under anti-PBC===============

%%==========Anti-periodic boundary condition==============
\textbf{Anti-periodic boundary condition:} %
We have also considered cylinders with anti-periodic boundary condition (ABC) in the $\hat{y}$ direction in order to test the extent to which our results are representative of the two-dimension limit. As shown in Fig.\ref{Fig:ABC} and previous studies\cite{Dodaro_PRB_2017}, the influence of changing boundary condition around the cylinder is significant. For example, the ground state of short cylinders with length $L_x\leq 48$, e.g., $L_x=32$ cylinder in Fig.\ref{Fig:ABC}, forms charge stripes of wavelength $\lambda=\frac{1}{\delta}$, which are completely filled with holes. However, this turns out to be a finite-size effect; the bulk of longer cylinders with length $L_x\geq 64$ %, e.g., $L_x=160$ cylinder in Fig.\ref{Fig:ABC}, 
 exhibits half-filled stripes with wavelength $\lambda=1/2\delta$, which is the same as the charge stripes of cylinders with periodic boundary conditions. Examples of the charge density distribution of cylinders with ABC in the $\hat{y}$ direction are given in Fig.\ref{Fig:ABC} for $\delta=12.5\%$. The ABC apparently affects the balance between filled and half-filled charge stripes; the former are stabilized in a finite region close to the open boundaries of the cylinder, while half filled stripes are robust in the bulk.

%=========Fig6: Entropy and central charge===========

%%========Central charge===========
\textbf{Central charge:} %
A key feature of a  Luther-Emery liquid is that it has a single gapless mode, i.e. it is expected to exhibit central charge $c=1$. The central charge can be obtained by calculating the von Neumann entropy $S=-\rm Tr \rho ln \rho$, where $\rho$ is the reduced density matrix of a subsystem with length $l$. For critical systems in 1+1 dimensions, it has been established\cite{Pasquale_JSMTE_2004} that 
$S(l)=\frac{c}{6}{\rm ln} (l)+\tilde{c}$, %for open systems,
 where $c$ is the central charge of the CFT and $\tilde{c}$ denotes a model dependent constant. For finite cylinders with length $L_x$, we can fix $l=\frac{L_x}{2}$ to extract the central charge $c$. Figure \ref{Fig:Entropy} shows the von Neumann entropy $S(\frac{L_x}{2})$ for different system sizes at different doping levels. The inset of the figure shows the fitted central charge $c$ as a function of doping level $\delta$. Although the extracted value of central charge $c$ is slightly larger than $c=1$, we suspect that this is within the uncertainty of the calculation.  The result is  roughly consistent with one gapless charge mode with $c=1$, which thus provides additional evidence for the presence of  a Luther-Emery liquid in the doped $t$-$J$ model.

%%============Summary and Discussion==============
\textbf{Summary and Discussion:} %
The presence of power-law superconducting correlations with $K_{sc}<2$ %close to 1
in a 4-leg ladder is an encouraging piece of evidence of the possible existence of a high temperature superconducting phase in 2D.  The $Q=4\pi\delta$ CDW correlations are reminiscent of the experimentally observed ``half-filled'' charge-stripe order that has been observed in previous DMRG studies of $t$-$J$ models,\cite{whitescal} in DQMC studies of the Hubbard model at elevated temperatures,\cite{edwin,Huang2} and experimentally in several cuprates.  The spin correlation length (shown in Fig. \ref{Fig:SDWP}) decreases monotonically with increasing $\delta$ from $\xi_s\simeq 6.5$ for $\delta=0$ to $\xi_s\simeq 4$ for $\delta = 12.5\%$.

It is still unclear  how the interplay between SC and CDW order should be expected to evolve with increasing cylinder circumference, $L_y$.  This uncertainty is exacerbated by the large number of nearly degenerate ground-state phases that were found previously\cite{Dodaro_PRB_2017} to be stabilized by relatively small changes in the microscopic parameters of the model.  The subtlety of the interplay between multiple phases is illustrated by changing the boundary conditions on the electronic wave-functions from periodic to anti-periodic. As shown in Fig. \ref{Fig:ABC},  on shorter cylinders (e.g. $L_x=32$), a distinct CDW state with $Q=2\pi\delta$ is stabilized. This state is reminiscent of the ``filled stripes'' found in Hartree-Fock calculations\cite{zaanen,machida,schulz} (where it is accompanied by long-range SDW order) and using various approximate methods\cite{simons} used in studies of the 2D Hubbard model.\footnote{Such a state was also found for 4-leg cylinders with $t^\prime \approx 0.06-0.08 t$ in Ref.\onlinecite{Dodaro_PRB_2017}.} In the present case, we find that while even for much longer flux-pierced cylinders, while the filled stripe state is observable locally for a finite region near the ends of the cylinders, far from the ends the CDW correlations have the same $Q=4\pi\delta$ ordering vector as in the fluxless cylinder.
  
One big question is the fate of   the magnetic correlations in the 2D limit.  For $\delta=0$, on theoretical grounds\cite{chakravfartyprl,charkravartyandgrevenprl} we know that $\xi_s$ should diverge with $L_y\rightarrow \infty$ since the ground-state of the spin 1/2 Heisenberg model is magnetically ordered in 2D.  The shorter correlation lengths of the doped systems suggests, but does not establish, that long-range antiferromagnetic order is unlikely to persist   in 2D for even  relatively modest values of $\delta$.

%%============Acknowledgement==============
\textbf{Acknowledgement:}
We would like to thank T. Devereaux, D. J. Scalapino, J. Tranquada, J. Zaanen, A. Broido,  Y. F. Jiang and J. Dodaro for insightful discussions. This work was supported by the Department of Energy, Office of Science, Basic Energy Sciences, Materials Sciences and Engineering Division, under Contract DE-AC02-76SF00515. Parts of the computing for this project was performed on the Sherlock cluster.

%%%%%%%%%%%%%%%%%%%%%%%%%%%%%%%%%
%\bibliography{Ref}

%\begin{thebibliography}{5}
%\end{thebibliography}
%\bibliographystyle{plain}
\bibliographystyle{ieeetr}
%\bibliography{bibs}

%%%%%%%%%%%%%%%%
\appendix 

\begin{center}
\noindent {\large {\bf Supplemental Material}}
\end{center}

\renewcommand{\thefigure}{S\arabic{figure}}
\setcounter{figure}{0}%reset counter
\renewcommand{\theequation}{S\arabic{equation}}%redefine command that creates equation no.
\setcounter{equation}{0}%reset counter

%==========FigS1: Finite-entanglement scaling===========
\begin{figure}
  \includegraphics[width=\linewidth]{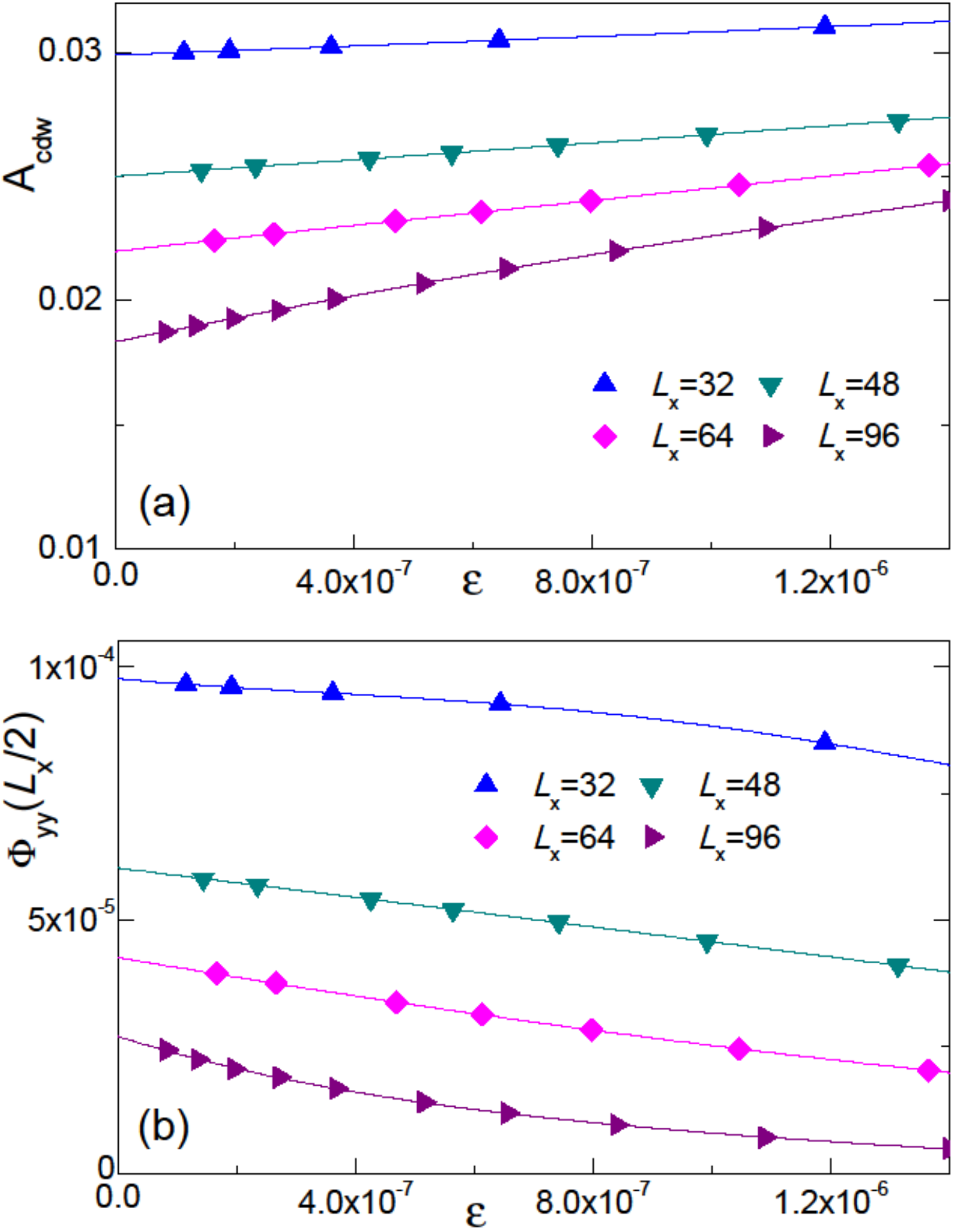}
  \caption{(Color online) (a) The amplitude of CDW $A_{cdw}$ and (b) superconducting correlation $\Phi_{yy}(\frac{L_x}{2})$ for $t$-$J$ model at doping level $\delta=12.5\%$ with $L_x=16\sim 96$. The solid lines denote fitting using quartic polynomial function.}\label{FigS:FES}
\end{figure}

%%=========Further calculation details=========
\section{I. Further calculational details}
To reliably describe the ground state properties, it is worth to mention that in the current study, we typically start our calculation with a random state. However, to elucidate the reliability of our results, we also double-check our calculation by adding a pinning field with the appropriate wavelength to stabilize the state. We find that in all the cases it is sufficient to add the pinning field during the initial sweeps of the calculation and ramp its amplitude to zero in a few subsequent sweeps. This happens only for the smallest number of states that we have considered, i.e., $m=2187$, while for the remaining calculation with $m>2187$ it is not necessary to hold a finite (even vanishingly small) pinning field to stabilize the stripe pattern. This gives us the same results as when we start from a completely random initial state without any pinning field, which undisputedly proves the reliability of our study regarding the charge density wave order.

Due to the presence of charge density oscillation in Fig.\ref{Fig:CDWP}(a), the superconducting correlations $\Phi_{\alpha\beta}(x)$ also exhibit similar spatial modulation with the charge density distribution (See Fig.\ref{FigS:SCCor}). This modulation, together with the significant boundary effect due to the presence of open ends of the cylinder, will make it very difficult, especially for short cylinders, to accurately determine the decaying behavior of the superconducting pair-field correlation function.%This could be one of the major reasons that previous DMRG studies\cite{Dodaro_PRB_2017} failed to find quasi-long range superconductivity.

In this work, we find a better way to determine the decaying behavior of superconducting correlation, which can minimize the effect of charge density modulation and the boundary effect simultaneously. As shown in the following, this allows us to directly and accurately determine the decaying behavior of the superconducting pair-field correlation $\Phi_{\alpha\beta}(x)$. For a given cylinder of length $L_x$, we calculate $\Phi_{\alpha\beta}(\frac{L_x}{2})$ with reference bond located at the peak position at $x_0\sim \frac{L_x}{4}$ of charge density distribution. This can be considered as the superconducting order parameter, i.e., $\Phi_{\alpha\beta}(\frac{L_x}{2})$, for a given finite system, which allows us to minimize both effects induced by open boundaries and charge density modulation as well as the finite-size effect simutaneously. Examples of CDW amplitude $A_{cdw}(L_x)$ and superconducting correlations $\Phi_{yy}(\frac{L_x}{2})$ are shown in Fig.\ref{FigS:FES} for $L_x=32\sim 96$ cylinders at doping level $\delta=12.5\%$. For a given $L_x$ cylinder, we extrapolate $A_{cdw}(L_x)$ and $\Phi_{yy}(\frac{L_x}{2})$ to the limit $\epsilon=0$ using quartic polynomial function. It is worth to mention that the linear regression $R^2$ in the fitting is always larger than $99.99\%$ and $99.97\%$ for $A_{cdw}(L_x)$ and $\Phi_{yy}(\frac{L_x}{2})$, respectively, suggesting the reliability of the extrapolation.

%==========FigS2: Ground state energy===========
\begin{figure}
  \includegraphics[width=\linewidth]{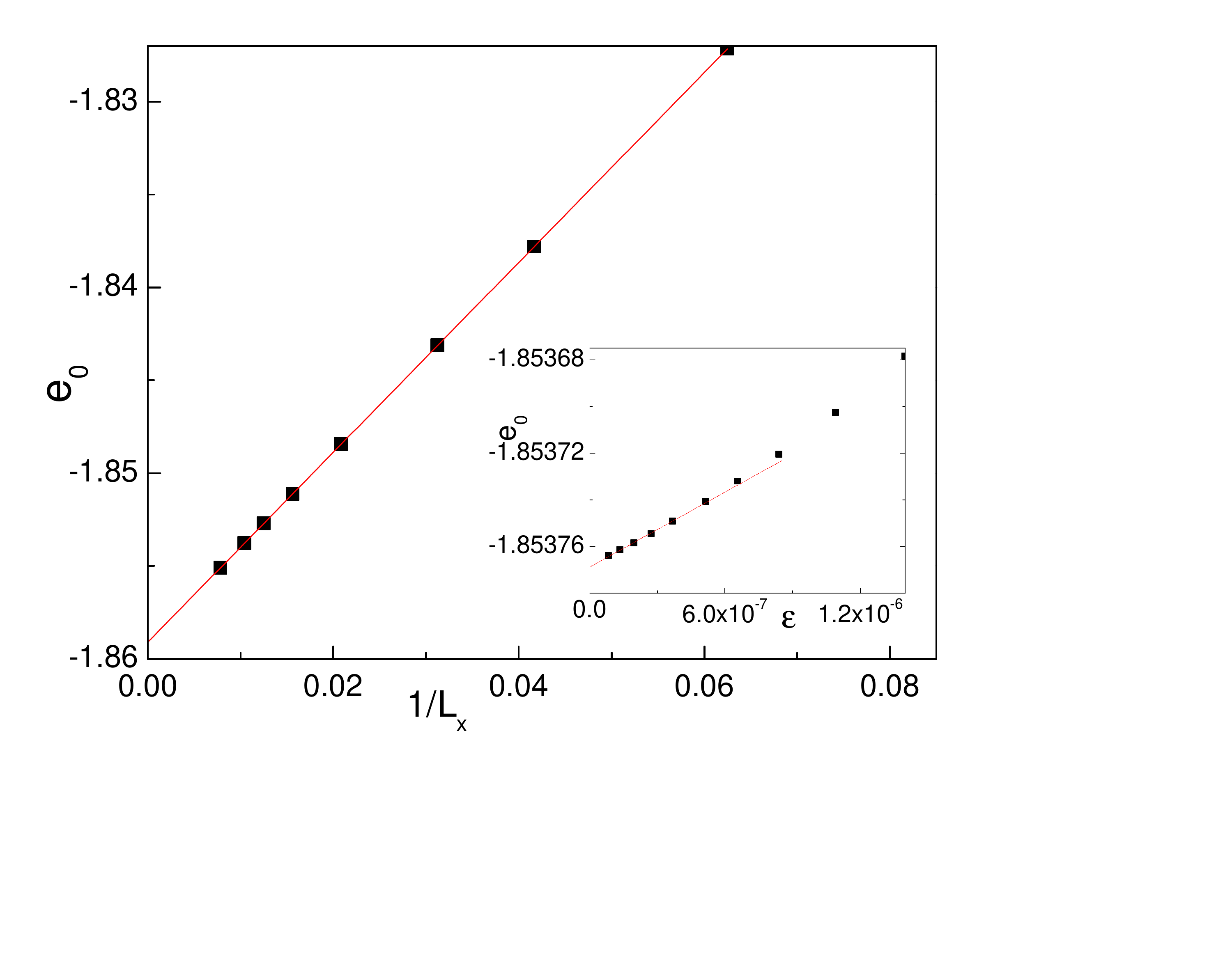}
  \caption{(Color online) Ground state energy persite $e_0$ for $t$-$J$ model at doping level $\delta=12.5\%$ as a function of the inverse cylinder length $L_x$. Inset: Example of truncation error $\epsilon$ extrapolation of $e_0$ for $L_x=96$ cylinder at doping level $\delta=12.5\%$.The red lines show the extrapolation using linear function. Here $t=3$ and $J=1$.}\label{FigS:EG}
\end{figure}

%%=========Ground state energy=========
\section{II. Ground state energy} %
In the inset of Fig.\ref{FigS:EG}, we show the example of truncation error $\epsilon$ extrapolation of the energy persite $e_0=E_0/N$, where $E_0$ is the total energy of a system with $N$ lattice sites, for $L_x$=96 cylinder at doping level $\delta=12.5\%$ with $t$=3 and $J$=1. By keeping $m=2187\sim 15000$ number of states, we are able to converge to the true ground state of the system by preserving all symmetries of the Hamiltonian, including the $SU(2)$ spin rotational symmetry, lattice translational symmetry in $\hat{y}$ direction and reflection symmetry in $\hat{x}$ direction. The truncaiton error extrapolation using linear function with $m=6000\sim 15000$ gives us $e_0=1.85377(1)$. The ground state energy $e_0$ of other cylinders at different doping levels can be obtained similarly. Finally, we can obtain acurate estimates of the ground state energies in the long cylinder length limit, i.e., $L_x=\infty$, by carrying out consective extrapolation in the truncation error, followed by the inverse cylinder length. The extrapolation to the limit $L_x=\infty$ for doping level $\delta=12.5\%$ is shown in Fig.\ref{FigS:EG}, in which all energies for cylinder lengths $L_x=16\sim 128$ fall perfectly onto the linear fit with  linear regression $R^2=1$. This gives us the energy $e_0=-1.85910(1)$ for doping level $\delta=12.5\%$, $e_0=-1.73072(1)$ for $\delta=10.0\%$, $e_0=-1.64213(1)$ for $\delta=8.33\%$ and $e_0=-1.46044(1)$ for $\delta=5.0\%$.

%=======FigS3: Superconducting pair-field correlation function===============
\begin{figure}
  \includegraphics[width=\linewidth]{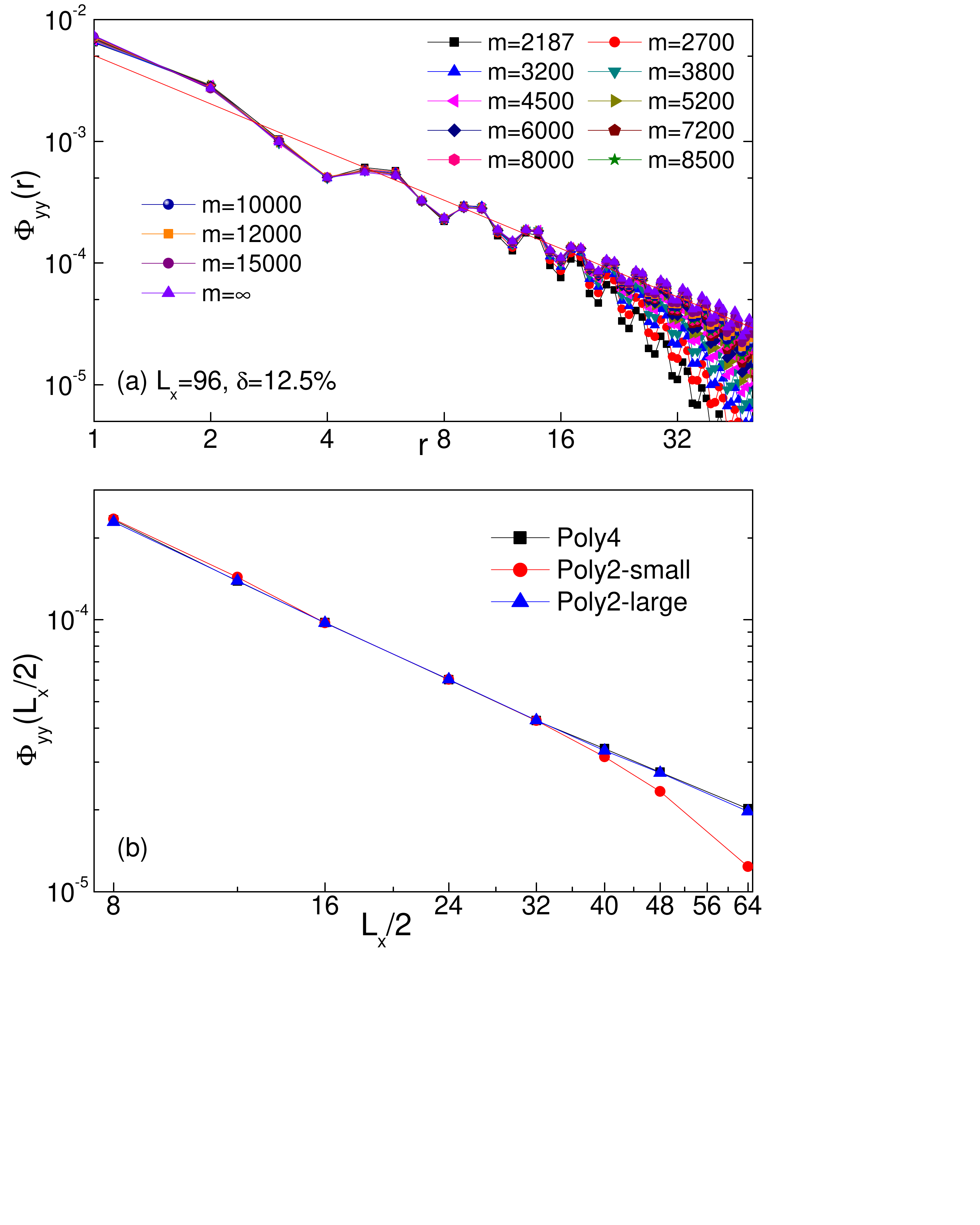}
  \caption{(Color online) (a) Superconducting pair-field correlation function $\Phi_{yy}(r)$ on $L_x=96$ cylinder at doping level $\delta=12.5\%$ by keeping $m=2187\sim 15000$ states, where $r$ is the distance between two Cooper pairs in $\hat{x}$ direction. (b) The extrapolated superconducting order parameter $\Phi_{yy}(\frac{L_x}{2})$ using quartic polynormail (Poly4) and quadratic polynomial (Poly2) function at doping level $\delta=12.5\%$ in the double-logarithemic plot (see text for detail). Here $L_x=16\sim 128$.}\label{FigS:SCCor}
\end{figure}

%=======FigS4: Exponents from Friedel oscillation and density-density correlation function========
\begin{figure}
  \includegraphics[width=\linewidth]{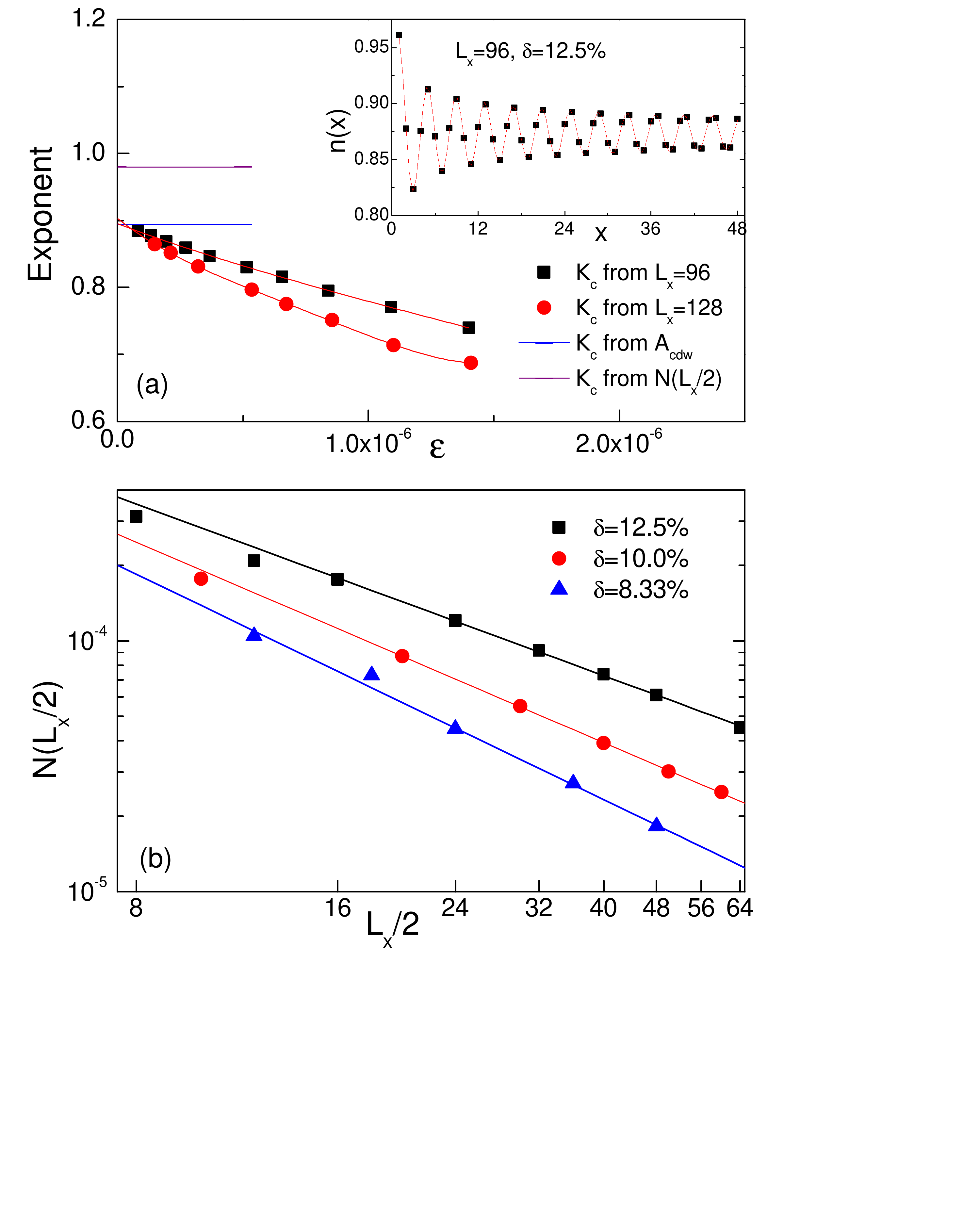}
  \caption{(Color online) (a) Luttinger exponent $K_c$ extracted from local density profile $n(x)$ via Friedel oscillations for $L_x=96$ and $L_x=128$ cylinders at doping level $\delta=12.5\%$, by keeping $m=2187\sim 15000$ number of states. The blue and purple lines represent the exponents determined from $A_{cdw}$ and density-density correlation function shown in (b). Inset: Fit of the local density profile $n(x)$ (solid line) on $L_x=96$ cylinder at doping level $\delta=12.5\%$ using function $n(x)=n_0 + \delta n\ast {\rm cos}(2k_F x + \phi)x^{-K_c/2}$, where $x=1\sim \frac{L_x}{2}$ is the rung index. (b) Finite-size scaling of extrapolated $N(\frac{L_x}{2})$ to the limit $m=\infty$ in a double-logarithmic plot at different doping levels.}\label{FigS:ExpK}
\end{figure}

%%=========Superconducting correlation function=============
\section{III. Superconducting correlation function}\label{SM:SCCor} %
Figure \ref{FigS:SCCor}(a) shows the superconducting pair-field correlation function $\Phi_{yy}(x)$ for $x=1\sim 50$ along $\hat{x}$ direction on the $L_x=96$ cylinder by keeping $m=2187\sim 15000$ number of states at doping level $\delta=12.5\%$. The purple triangle labels the extrapolated value to the limit $\epsilon=0$, i.e., $m=\infty$, using quartic polynomial function, which is consistent with the power-law decaying behavior $\Phi_{yy}(x)\propto |x|^{-K_{sc}}$ as indicated the red solid line.

Figure \ref{FigS:SCCor}(b) plots the extrapolated superconducting order parameter $\Phi_{yy}(\frac{L_x}{2})$ on cylinders with length $L_x=16\sim 128$ at doping level $\delta=12.5\%$ fitted by different orders of polynomial function. The black squares label the extrapolated $\Phi_{yy}(\frac{L_x}{2})$ using quartic polynomial function (Poly4). The red circles denote the results fitted by quadratic polynomial function by keeping up to $m=8000$ number of states (Poly2-small). On the contrary, the blue triangles represent results fitted by the same quadratic polynomial function but only using 5 data points with largest number of states (Poly2-large) for each cylinder, e.g., $m=8000\sim 15000$ for $L_x=96$ cylinder. From the figure we can clearly see that both Poly4 and Poly2-large fittings are consistent with each and enough to capture the long distance behavior of the superconducting pair-field correlation, while the Poly2-small fitting by keeping up to $m=8000$ is not. This may explain the absence of long-range superconductivity in previous studies since the numerical accuracy or the number of states is not high enough.

%=======Friedel oscillation of density profile and density-density correlation function===============
\section{IV. Friedel oscillation of density profile and density-density correlation function}\label{SM:DensityCor} %
One of the most reliable estimation of the Luttinger exponent $K_c$ in DMRG simulation is based on the Friedel oscillations of the density profile induced by the open boundaries of cylinder\cite{White_PRB_2002}. In this work, we use $n(x)=n_0 + \delta n\ast {\rm cos}(2k_F x + \phi) x^{-K_c/2}$ to fit the local density profile to extract the Luttinger exponent $K_\delta$ through Friedel oscillations. Here, $\delta n$ is the nonuniversal amplitude, $\phi$ is a phase shift, $n_0$ is the background density and $k_F$ is the Fermi wavevector. Example of the fitting is given in the inset of Fig.\ref{FigS:ExpK}(a) for $L_x=96$ cylinder at doping level $\delta=12.5\%$ with rung index $x=1\sim \frac{L_x}{2}$ by keeping $m=15000$ number of states. The main panel shows the extracted Luttinger exponent $K_c$ for both $L_x=96$ and $L_x=128$ cylinders at the same doping level. In the limit of $m=\infty$, the extracted exponent for both cylinders is consistent with that determined from the finite-size scaling of charge density wave order parameter $A_{cdw}(L_x)$ (see Fig.2 in the main text).

For comparison, we also show the exponent extracted from finite-size scaling of the charge density-density correlation function, which is defined as $N(r)=\langle (\hat{n}(x_0)-\langle \hat{n}(x_0)\rangle)(\hat{n}(x_0+r)-\langle \hat{n}(x_0+r)\rangle)\rangle$. Here $x_0$ labels the rung index of the reference site given in Fig.1. Figure \ref{FigS:ExpK}(b) shows the finite-size scaling of the extrapolated $N(\frac{L_x}{2})$ at different doping levels. Similar with both $A_{cdw}(\frac{L_x}{2})$ and $\Phi_{yy}(\frac{L_x}{2})$ given in the main text, $N(\frac{L_x}{2})$ also decays with a power-law whose exponent $K_c$ can be extracted by fitting the results using $N(\frac{L_x}{2})\propto (\frac{L_x}{2})^{-K_c}$. Although the exponent determined from $N(\frac{L_x}{2})$ is slightly larger (around $10\%$) than that extracted from $A_{cdw}(L_x)$ and Friedel oscillations of density profile, they are all qualitatively consistent.

\end{document}